\documentclass{aa}    
\usepackage{psfig}

\def\ros{{\sl ROSAT }}

\def\ein{{\sl Einstein}}
\def\exo{{\sl EXOSAT }} 
\def\G{$\Gamma_{\rm x}$ }

\def\farcs{\hbox{$.\!\!^{\prime\prime}$}}  
\def\approxlt{\mathrel{\hbox{\rlap{\lower.55ex \hbox {$\sim$}}
        \kern-.3em \raise.4ex \hbox{$<$}}}}
\def\approxgt{\mathrel{\hbox{\rlap{\lower.55ex \hbox {$\sim$}}
        \kern-.3em \raise.4ex \hbox{$>$}}}}
\begin{document}

  \thesaurus{03         
              (13.25.2;  
               11.05.1;  
               11.09.1;  
               11.03.4;  
               12.04.1)} 
   \title{X-ray study of the NGC\,383 group of galaxies \\
          {\large{\bf and the source 1E\,0104+3153}}}
   \author{Stefanie Komossa \and Hans B\"ohringer}  
\offprints{St. Komossa, \\
              skomossa@xray.mpe.mpg.de}
\institute{
       Max-Planck-Institut f\"ur extraterrestrische Physik,
         Postfach 1603, D-85740 Garching, Germany
            }
\date{Received: October 1998; accepted: 11 January 1999}
   \maketitle
\markboth{St. Komossa \& H. B\"ohringer: X-ray study of the NGC\,383 group and 1E\,0104+3153}
{St. Komossa \& H. B\"ohringer: X-ray study of the NGC\,383 group and 1E0104+3153}
   \begin{abstract}
We present results from an analysis of the
X-ray properties of the NGC\,383 galaxy group
based on \ros PSPC and HRI data.
X-ray emission can be traced out to $\sim 1 h_{50}^{-1}$ Mpc,
the estimated virial radius of the system.
We determine a total mass of $6\,10^{13} h_{50}^{-1}$
M$_{\odot}$ for the group inside this radius
with a gas mass fraction of 21\%. The intragroup
gas temperature of $1.5$ keV is both consistent
with the galaxy velocity dispersion and the
X-ray luminosity - temperature relation of groups
and clusters suggesting that the group is
fairly relaxed. This is also indicated 
by the almost spherically symmetric appearance
of the group's X-ray halo.

The X-ray properties of the radio galaxy NGC\,383 (3C\,31) which is located
near the center of the group are discussed.
Its spectrum is best described by a two-component model,
consisting of emission from a low-temperature Raymond-Smith plasma, and a hard tail.
The emission from NGC\,383 is not resolved by
the \ros HRI. 
The possible interaction of the radio jets of 3C\,31 with the IGM is studied.

A spatial, spectral and temporal analysis of the 
\ein~source 1E0104+3153 located within the field of view is 
performed, one goal being the identification of the optical
counterpart (with both, a high-redshift BAL quasar and
a nearby elliptical galaxy, member of a small group, located
within the \ein~X-ray error circle).
We find evidence that the IGM of the small group contributes
significantly to the X-ray emission of 1E0104, which
can be described by a Raymond-Smith model of $kT \simeq$ 2 keV 
and a soft X-ray luminosity of $L_{\rm x} \simeq 3\,10^{43}$ erg/s.

   \end{abstract}
%
\section{Introduction}

NGC\,383, of type S0 (Arp 1968), is a member of a bright chain of galaxies 
(Arp 1966, Arp \& Bertola 1971, Kormendy \& Bahcall 1974),
which itself belongs to a group of galaxies (Zwicky et al. 1961) 
 located within the Perseus-Pisces filament. 
NGC\,383 is at redshift $z$=0.017 and has a companion galaxy in 33\arcsec~distance. 

NGC\,383 is a moderately bright radio galaxy (3C\,31) and has been extensively 
studied at radio wavelengths in the past (e.g., Macdonald et al. 1968, Burch 1977,1979,
Klein \& Wielebinski 1979, Fomalont et al. 1980, Ekers et al. 1981,
van Breugel 1982, Strom et al. 1983, Condon et al. 1991, 
Andernach et al. 1992, Artyukh et al. 1994,
Lara et al. 1997, Henkel et al. 1998).
It shows a symmetric edge darkened   
double-source structure with two strong jets.
The origin of the 
structures seen in the radio morphology of 
dominant cluster/group members is still not well understood. 
The structures of the jets of NGC\,383 were interpreted
by Blandford \& Icke (1978) as due to tidal interaction with the companion
galaxy NGC\,382. However, Fraix-Burnet et al. (1991) found no evidence 
for interaction between the two galaxies. 
Further jet models were presented by Bridle et al. (1980) and Bicknell (1984). 
Whereas Butcher et al. (1980) reported evidence for a detection of the jet
at optical wavelengths, this was not confirmed by later studies (Keel 1988, Owen et al. 1990,
Fraix-Burnet et al. 1991).

The optical spectrum of the nucleus is characterized by emission lines. 
The optical emission was found to be extended, and in form of a rotating 
disk (Owen et al. 1990) 
or ring (Fraix-Burnet et al. 1991) that coincides with a dust-ring reported 
by Butcher et al. (1980).  
The emission line ratios in the disk were found to be similar to those 
at the nucleus  
and Owen et al. concluded that the disk ionization is probably driven
by the nucleus.

NGC\,383 is the brightest member of a rich group of galaxies. 
The group is included in the Zwicky catalogue; membership studies based on velocity measurements
were performed by Moss \& Dickens (1977), Garcia (1993), Sakai et al. (1994), and 
Ledlow et al. (1996), who included 25, 22, 16, 32 member galaxies, respectively.  
Among the 27 groups studied, Ledlow et al. found 
the galaxy velocity distribution
to significantly deviate from a Gaussian in 55\% of them; 
the NGC\,383 group was not among those.
Sakai et al., using a standard estimator as in Heisler et al. (1985),
derived a virial mass of $M_{\rm vir}$ = 0.2\,10$^{14}$ $h M_{\odot}$, 
and a mass-light ratio $M/L$ = 240.
Extended X-ray emission from the group was first detected by \ein~
(Fabbiano et al. 1984).  The data were 
analyzed by Morganti et al. (1988) in the course of a 
large sample study 
of the effect of gas pressure on radio sources.

To perform 
a detailed spatial and spectral 
analysis of
the X-ray emission from NGC\,383, the intra-group medium (IGM), and an investigation
of the relation with
the radio jet, a \ros PSPC observation was applied for.
These data presented here have been previously partly analyzed by
Trussoni et al. (1997; T97 hereafter) in a study of hot coronae in nearby
radio galaxies. Here, we extent their analysis and focus on the properties of 
the IGM and the NGC\,383 group as a whole, in particular, the determination
of gas and total mass; and the nature of the X-ray emission
from NGC\,383.  
We also briefly discuss new HRI data which were retrieved from the archive. 

Also located in the field of view is the high-redshift ($z$=2.027)
BAL quasar QSO 0104+3153. X-ray emission from the direction of this quasar was
discovered in the \ein~Medium Sensitivity Survey (Stocke et al. 1984). 
The QSO, and a nearby (10\arcsec)
giant elliptical galaxy, member of a small group, were identified as
possible counterparts of the X-ray emission. Due to its proximity to
the elliptical, the quasar is considered as prime candidate for
lensing by halo stars. Optical variability of $\sim$0.5$^{\rm m}$ 
was detected by Gioia et al. (1986). The \ros PSPC spectrum
was briefly discussed in a large sample study by Ciliegi \& Maccacaro (1996).
The identification
of the source (QSO, elliptical galaxy or IGM of the small group) remained unclear.
Here, we perform a detailed study of the properties of this X-ray source
and address the counterpart question
on the basis of (i) long- and short-term variability behaviour,
(ii) X-ray spectral shape, (iii) improved spatial position obtained with the HRI.

Physical parameters are calculated for $H_{\rm o}$ = 50 km/s/Mpc, 
$q_{\rm o}$ = 0.5 and assuming
the galaxies/group to follow the Hubble flow.
For the distance of the NGC\,383 group, 1\arcsec~corresponds to a scale of 0.5 kpc.   

\section{X-ray data}             

\subsection{PSPC} 

The group was the target of a pointed \ros 
PSPC (Tr\"umper 1983; Pfeffermann et al. 1987) observation with NGC\,383 in the
center of the field of view. 
Our observation was performed from July 28 -- 29, 1991 with a duration
of 27.5 ksec. The satellite was not in wobble mode during this observation. 

The background was usually determined in a source-free ring around
the target source (for details see below). 
The data were corrected for vignetting
and dead-time, using the EXSAS software package (Zimmermann et al. 1994).
Widely extended X-ray emission is present (Fig. 1). The central region 
appears roughly circularly symmetric.
Emission from the group can be traced out to a distance of $\sim$33\arcmin~radius
($\sim$ 1 Mpc).
The total countrate (channels 52--201) within this region is 0.58 $\pm{0.01}$ cts/s.
Several group ellipticals are individually detected,
the brightest with countrates of 0.024$\pm{0.001}$ cts/s (NGC\,383), 
0.0058$\pm{0.0006}$ cts/s (NGC\,379), 0.0124$\pm{0.0008}$ cts/s (NGC\,380),
0.0041$\pm{0.0005}$ cts/s (NGC\,385), and 0.0021$\pm{0.0006}$ cts/s (NGC\,384)
as summarized in Table \ref{fitres2}.
X-ray emission from the direction of 1E\,0104 is detected 
with 0.028$\pm{0.001}$ cts/s.  

To carry out the spectral analysis of the individual sources,
photons in
the amplitude channels 11-240 were binned
according to a constant signal/noise ratio of $\ge$ 5$\sigma$.

Results given below refer to the PSPC data if not mentioned otherwise.

\begin{figure}   
\vbox{\psfig{figure=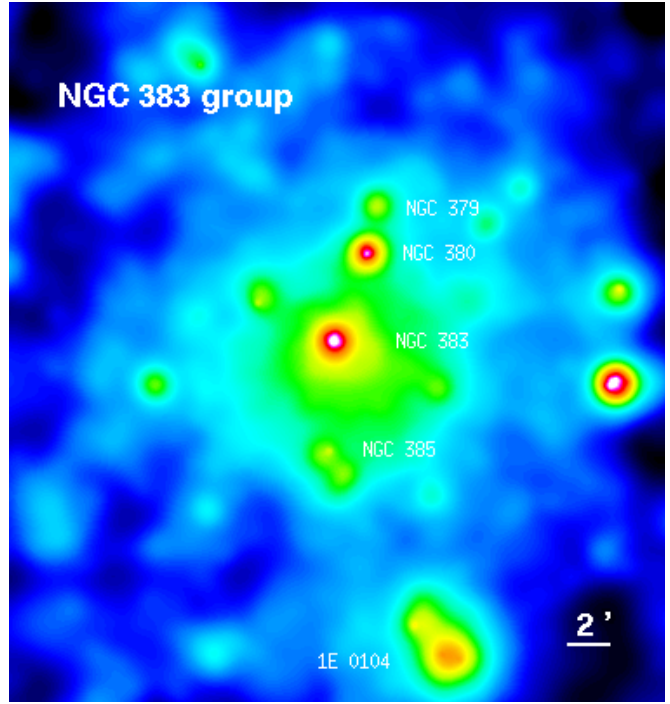,width=8.7cm,
bbllx=5.1cm,bblly=12.0cm,bburx=16.5cm,bbury=24.3cm,clip=}}\par
\caption[over0]{\ros PSPC X-ray image of 
the central region of the NGC\,383 group of galaxies. 
The image was smoothed with a variable 
Gaussian filter of widths $\sigma$ = 1\arcmin, 0.5\arcmin, 20\arcsec, 7.5\arcsec,
5\arcsec~and 3.5\arcsec.  
Several galaxies of the optical chain Arp\,331, and the \ein~source
1E\,0104+3153 are marked. Note the detection of another X-ray source
close to 1E\,0104.}
\label{over0}
\end{figure}

\subsection{HRI}

An HRI observation centered on NGC\,383 was performed between 
July 11 and 25, 1994 with a duration of 25.0 ksec. 
The data were retrieved from the archive.
(An earlier HRI observation, of Jan. 1992 with 2.8 ksec duration, is briefly
discussed in T97.) 
Due to the lower sensitivity of the HRI, essentially only the bright point-like sources
are detected. These data are analyzed mainly to study the spatial
structure of the X-ray emission from the bright group galaxies, and to 
search for variability 
in individual sources.  

\section{Group analysis}

\subsection{Spectral properties}

A Raymond-Smith (rs) model was fit to the extended X-ray emission.
If not noted otherwise, abundances were fixed to 0.35$\times$solar (Anders \& Grevesse 1989). 
In a first step, the group photons
within a circular region of radius 1000\arcsec~ were selected; in a second step, this region was split
into an inner circle (radius 200\arcsec), an intermediate ring (200\arcsec~out to 500\arcsec)
and an outer ring (500\arcsec~to 1000\arcsec). Point-like sources
were removed. 
To this end, we first carried out a source detection using the EXSAS software.
This resulted in about 35 sources detected within the total
field of view. For the brightest sources, extents were then determined 
by inspecting the radial source profiles (the radius where the source
profile merges with the local background taken as source extent) and by
comparison with the profile expected from the PSF of the detector (for a more
detailed discussion of NGC\,383 see Sect. 4.1.2). This resulted in source 
extraction radii of $\sim$100\arcsec~for the brightest central sources. 
The detected sources were then removed within circular regions centered on their
X-ray positions.  
As background we selected a source-free circular region in the
outer part of the fov. Since the spectrum below 0.5 keV is strongly dominated
by background, we usually excluded photons below this energy from the spectral
fit. 
Treating the cold absorption as free parameter always resulted in a value
(nearly) consistent
with the Galactic absorption towards
NGC\,383 ($N_{\rm gal} = 0.523\,10^{21}$ cm$^{-3}$, Dickey \& Lockman 1990) within
the errors. We therefore fixed $N_{\rm H}$ = $N_{\rm gal}$.

For the total emission within 1000\arcsec~ we find a temperature
$kT$ = 1.5$\pm{0.1}$ keV. The temperature values
derived for the three separate regions are consistent with this value and with each other
within the errors (Table \ref{fitres}).
This value of $T$ is somewhat lower than the one given in
Morganti et al. (1988; their Tab. 2) on the basis of an \ein~ IPC observation;
they estimate $kT \approx$ 3 keV. 
To check the robustness of the obtained temperature, we performed a few tests:
If we do {\em not} remove the point sources and re-fit the total spectrum
we get $kT$ = 1.4$\pm{0.1}$ keV; if we fit a thermal bremsstrahlung model
we obtain $kT$ = 1.4$\pm{0.2}$ keV; 
if an rs plasma with solar instead of depleted abundances is assumed, 
the quality of the fit slightly improves 
without affecting the value of the temperature within the errors and 
we derive $kT$ = 1.6$\pm{0.1}$ keV.
T97 obtain $kT$ = 1.5 keV for abundances of 0.35$\times$solar.   
Results of our spectral fits are summarized in Table \ref{fitres}.  

Using $kT$ = 1.5 keV and $N_{\rm H} = N_{\rm gal}$, 
the total (0.1--2.4 keV) luminosity within 33\arcmin~is 
$L_{\rm x} = 1.5\,10^{43}$ erg/s (= 1.3\,10$^{43}$ erg/s if the abundances
are fixed to the solar value). 

\subsection{Spatial analysis}

All bright ellipticals of the optical chain Arp\,331 are individually
detected in X-rays. 
The center of the X-ray emission maximum from NGC\,383
at $\alpha$ = 1$^{\rm h}$7$^{\rm m}$25.9, $\delta$ = 32\degr 24\arcmin 44\farcs5 (J\,2000)
coincides well with the position of the optical nucleus at
$\alpha$ = 1$^{\rm h}$7$^{\rm m}$25.0,
$\delta$ = 32\degr 24\arcmin 44\farcs8. 
For an overlay of the PSPC X-ray contours on an optical image of the NGC\,383 group
see T97 (their Fig. 1).

To derive physical properties of the group, we first assume spherical
symmetry of the extended X-ray emission, the emission to be 
centered on NGC\,383, and rough isothermality to hold.
For critical comments on and justification of `standard' assumptions see, e.g., 
B\"ohringer et al. 1998 (their Sects. 2, 3). 
A $\beta$-model (e.g., Cavaliere \& Fusco-Femiano 1976, 
Gorenstein et al. 1978, Jones \& Forman 1984)      
of the form 
\begin{equation} 
S = S_0 (1+ {r^2 \over {r_{\rm c}^2}})^{-3\beta + {1 \over 2}}  
\end{equation} 
was fit to the azimuthally averaged 
surface brightness profile of the PSPC observation (detected point sources were, again,
removed except emission from NGC\,383, inner bins were then excluded
from the fit since they are dominated by emission from NGC\,383 itself). 
This yields a central surface brightness $S_{\rm 0} = 2.79\,10^{-3}$ cts/s/arcmin$^2$, 
a slope parameter $\beta$ = 0.34 and a core radius $r_{\rm c}$ = 64 kpc.  
The gas mass enclosed inside 1 Mpc amounts to $M_{\rm gas}$ = 1.5\,10$^{13}$ M$_{\odot}$. 
Inspection of Fig. 1       
shows that NGC\,383 is not located perfectly in the center 
of the large scale X-ray emission, but offset by about 1\arcmin~to the North-East. 
In fact, using the best-fit beta-model to construct a synthetic model image
and subtracting the model image from the observed one 
leaves some residual extended emission to the SW and some `negative' emission levels
to the NE.   

\begin{figure}
      \vbox{\psfig{figure=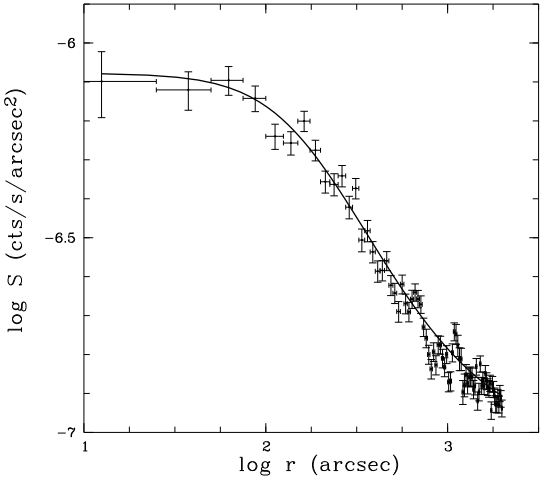,width=7.7cm,%
          bbllx=2.3cm,bblly=1.1cm,bburx=15.6cm,bbury=12.2cm,clip=}}\par
 \caption[profile]{Observed X-ray surface brightness profile (crosses) of the
NGC\,383 group of galaxies and 
best-fit $\beta$-modell (solid line). Due to the large extent of the emission,
the data have not been PSF-convolved (a test with a fit of a convolved model
did not show a significant change). Channels 52-201 were used to derive the 
surface brightness profile.  
}
 \label{profile}
\end{figure}

Therefore, in a second step, the surface brightness profile was centered
at $\alpha$ = 1$^{\rm h}$7$^{\rm m}$22.8, $\delta$ = 32\degr 23\arcmin 45\farcs7 
and re-fit after emission from NGC\,383 was removed within a segment.
In this case, we obtain $S_{\rm 0} = 2.67\,10^{-3}$ cts/s/arcmin$^2$, 
$\beta = 0.38\pm{0.03}$, and $r_{\rm c} = 73^{+18}_{-15}$ kpc (errors are 1$\sigma$).
The only slight change of the fit parameters underlines the robustness of the fit.
The beta-model fit to the surface brightness profile is shown in Fig. \ref{profile}.
Again constructing a model image and subtracting this from the observed
one now does not show any residual emission except the point-like sources. 
This model is also used for the gas and total mass estimate below. 

These results differ from the previous analysis of T97 in yielding a smaller
core radius (they derived $r_{\rm c}$ = 230 kpc)
and shallower slope 
(they found $\beta$ = 0.6) for the data presented here. 
We have checked that our results are robust ($\beta$ varying 
between $\sim$ 0.3 and 0.45 in extreme cases) against non-removal 
of individual sources (up to all), shifts of the central position, 
and to systematic exclusion of individual bins (e.g., the inner or outer ones)
before fitting the surface brightness profile.
Our best fit leads to a larger ratio of gas to total mass at large radii.  
It also has consequences when comparing the pressure of the
IGM with the non-thermal pressure of the radio jet (Sect. 5.1.3).
 
  \begin{figure}
      \vbox{\psfig{figure=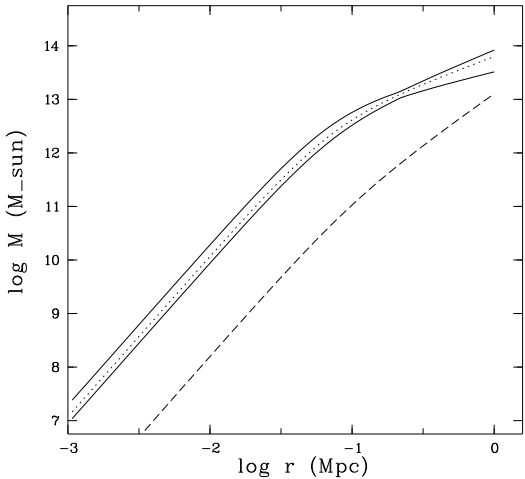,width=7.7cm,%
          bbllx=2.3cm,bblly=1.1cm,bburx=15.6cm,bbury=12.2cm,clip=}}\par
 \caption[mass]{Radial mass profile for the NGC\,383 group of galaxies.  
The dashed line gives the profile of gas mass as obtained from the best-fit
$\beta$-modell, the solid lines correspond to the profile of the total
mass as described in the text, and the dotted line corresponds to the 
isothermal case. 
}
 \label{mass}
\end{figure}

\subsection{Mass determination}

For the beta-modell applied in the previous section, the gas 
density distribution is given by 
\begin{equation}
n_{\rm gas} = n_0 (1+ {r^2 \over {r_{\rm c}^2}})^{-{3\over 2}\beta} ~~.
\end{equation}
This implies a central density $n_0$ = 1.3\,10$^{-3}$ cm$^{-3}$ and a gas mass within 1\,Mpc of
$M_{\rm gas}$ = 1.3\,10$^{13}$ M$_{\odot}$. 

Assuming spherical symmetry and the group to be approximately in hydrostatic
equilibrium, the total gravitating mass follows the relation 
\begin{equation}
M_{\rm total}(r) = -{k \over {\mu m_{\rm p} G}}\,T(r)\,r\,({r \over T}{dT \over dr} +
{r \over \rho}{d\rho \over dr})~~. 
\end{equation}   
With the observed parameters this results in an integrated total mass of 
$M_{\rm total}$ = 6.3\,10$^{13}$ M$_{\odot}$ within 1 Mpc radius 
and a gas mass fraction of 21\%.
Using instead the galaxy velocity dispersion as derived from optical
observations, $\sigma$ = 466 km/s (Ledlow et al. 1996; see also 
Sakai et al. 1994), and a core
radius of 73 kpc we get a mass of $\sim$7\,10$^{13}$ M$_{\odot}$ 
within 1 Mpc, which is well consistent with the value obtained purely
from the X-ray data. 

The profile of total and gas mass is displayed in Fig. \ref{mass}.
Errors on the mass $M_{\rm total}$ are obtained from the temperature
range allowed by the X-ray spectral analysis, $kT$
= 1.5$\pm{0.2}$ keV,   
and a temperature profile for a family of $\gamma$ models with 
polytropic index $\gamma$ in the range 0.9 -- 1.3. 
In the polytropic models the nominal temperature is fixed at the core radius. 

\subsection{Radio -- X-ray morphology}

\begin{figure} 
\psfig{file=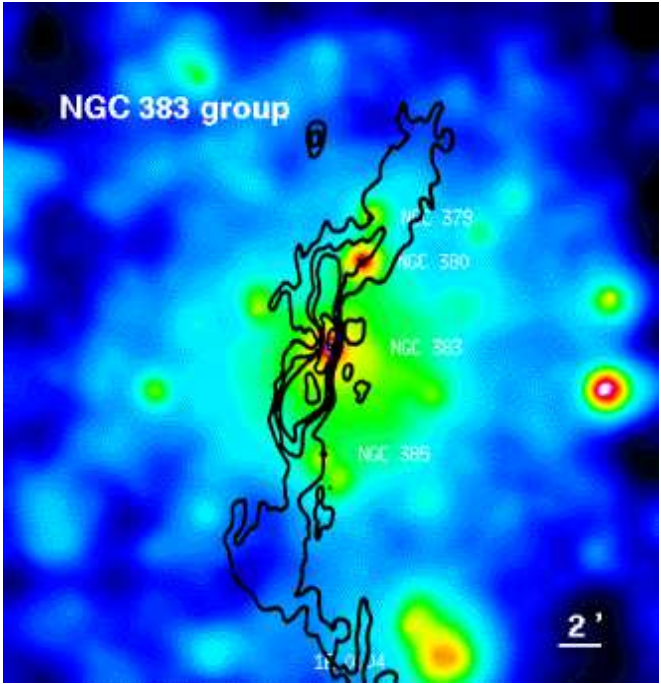,width=8.7cm}
\caption[over_rx]{Same as Fig. 1, but with overlay of the radio jet contours 
(Strom et al. 1983, 0.6 GHz map) on the
X-ray image.  
}
\label{over_rx}
\end{figure}

There are some spectacular examples of pressure interaction between the radio 
and X-ray gas in clusters of galaxies (e.g., B\"ohringer et al. 1993, 1995, 
Harris et al. 1994, 
Clarke et al. 1997, Otani et al. 1998).     
In the present case, we do not find conspicuous morphological 
correlations between radio- and X-ray emission (Fig. \ref{over_rx}).
This may be partly due to the narrowness of the jet, the still limited spatial
resolution of the \ros PSPC, and the 2D view of the 3D source structure.

Changes of the jet orientation angle near the locations of some optical chain galaxies,
now also detected as strong X-ray sources,  
were already noted by Strom et al. (1983).  

\section{Individual sources}

\subsection{NGC\,383 = 3C\,31} 

\subsubsection{Spectral analysis}

As background we chose (i) a source-free ring around the target source,
and (ii) a source-free circular region near the target source. This allows
to check for sensitivity against the background correction. All majour fits
were repeated for both background geometries. Further, we note that the source emission 
in the first bin (below 0.4 keV) is weak. We repeated all fits after having removed
the first bin from the spectrum. In all cases, the results presented below are 
found to be robust. 

First, several single component models were fit to the X-ray
spectrum, starting with a powerlaw (pl). 
Although this model
fits the X-ray spectrum ($\chi{^{2}}_{\rm red}$ = 1.3), 
the derived parameters 
are unusual. The slope is extremely steep \G $\simeq$ --5, and there is evidence
for strong excess absorption (about 8 times the Galactic value).  
A single steep pl may be mimicked by a flat pl plus soft excess.
Parameterizing the excess as black body and fixing \G=--1.9, we do not
find a successful fit. This also holds for a single pl in which absorption
is fixed to the Galactic value ($\chi{^{2}}_{\rm red}$ = 3.3).  
A single rs model with metal abundances of 0.35 $\times$ solar 
does not give an acceptable fit, either ($\chi{^{2}}_{\rm red}$ = 2.4).
Lowering the abundances up to $\approxlt$ 0.1 $\times$ solar 
yields an acceptable fit (see also T97), but such low abundances are unexpected
for dominant group galaxies{\footnote{E.g., Fabbiano et al. (1994)
find that assuming typical stellar mass loss rates, 
originally primordial
gas would be 10\% metal-enriched after 10$^7$ yr; 
see also Sarazin (1997). \\ 
For a recent thorough discussion of the issue of single-$T$ models
of very subsolar abundances vs. two-component models of $\sim$solar abundances
see also Buote \& Fabian (1998).}}.
 
In a second step, two-component models consisting of contributions
from both, a rs plasma and a pl source, or two rs sources were applied.    
In the rs+pl description, the pl index was fixed to \G=--1.9 
(the value typically observed in Sy; since we want to specifically test
for the presence of an AGN). This provides a good 
fit ($\chi{^{2}}_{\rm red}$ = 1.1). If $N_{\rm H}$ is treated as additional
free parameter, it is found to be of the order of the Galactic
value. For the rs component, we obtain $kT \simeq$ 0.6 keV (error contours 
are displayed in Fig. \ref{chi2}). The pl component contributes with a 
(0.1--2.4 keV) luminosity of 
$L_{\rm x,pl} = 3.5\,10^{41}$ erg/s, the total (0.1--2.4 keV) luminosity is 
$L_{\rm x} = 4.7\,10^{41}$ erg/s (for $N_{\rm H}$ fixed to $N_{\rm gal}$).   
Alternatively, the spectrum can be fit with a double rs model.
In this case we find the hotter component to be ill-constrained. 
If we fix $kT_2$ = 1.5 keV, we get $kT_1$ = 0.4$\pm{0.2}$ keV 
($\chi{^{2}}_{\rm red}$ = 1.1; we used $N_{\rm H} = N_{\rm gal}$)
and luminosities of $L_{\rm x,T_1} = 1.7\,10^{41}$ erg/s, 
$L_{\rm x,T_2} = 2.9\,10^{41}$ erg/s.
For comparison, if $kT_2$ = 5 keV is chosen, $kT_1$ = 0.5$\pm{0.2}$ keV
($\chi{^{2}}_{\rm red}$ = 1.0) and $L_{\rm x,T_1} = 2.0\,10^{41}$ erg/s, 
$L_{\rm x,T_2} = 2.7\,10^{41}$ erg/s.  
The results of the spectral fits are summarized in Table \ref{fitres}.

  \begin{figure}[thbp]
      \vbox{\psfig{figure=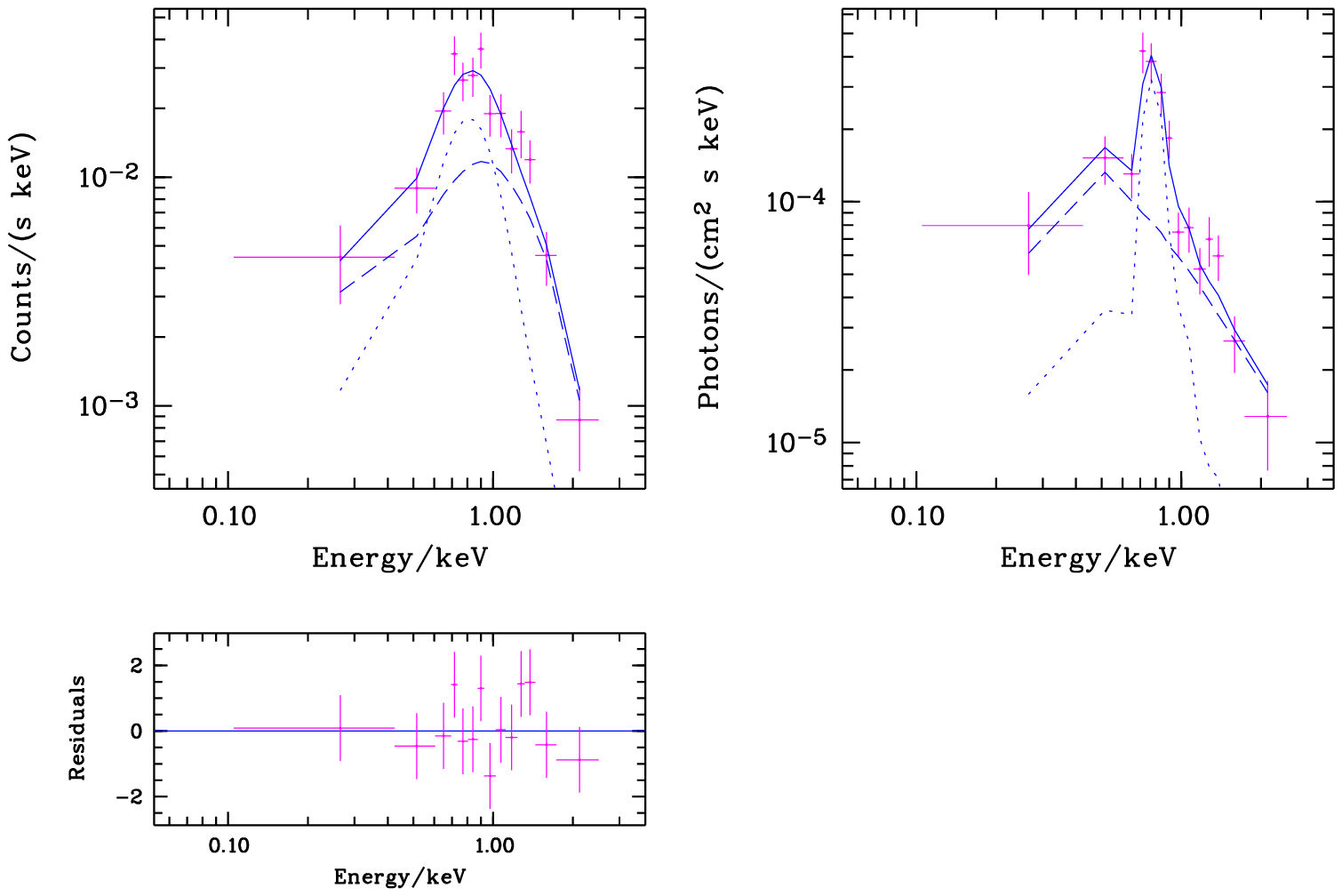,width=7.3cm,%
          bbllx=2.5cm,bblly=1.1cm,bburx=10.1cm,bbury=11.7cm,clip=}}\par
\caption[SEDx]{The first panel shows the observed X-ray spectrum of 
NGC\,383 (crosses)
and the best-fit rs+pl model (solid line). 
The second panel displays the residuals for this
model.} 
\label{SEDx}
\end{figure}

\begin{table*}             
     \caption{Spectral fits to individual galaxies of the group  
(NGC\,383, NGC\,380 and NGC\,379) and the intra-group medium (IGM)
emission 
(pl = power law,
rs = Raymond-Smith model). 
$T$ = temperature of 
rs component, $N_{\rm H}$ = cold absorbing column, $Z$ = metal abundances, relative
to solar (Anders \& Grevesse 1989),
d.o.f. = degrees of freedom.
The error contours for the rs+pl model of NGC\,383, and the rs model
of the total IGM emission, are displayed in Fig. \ref{chi2} and \ref{chi}, respectively.
              }
     \label{fitres}
      \begin{tabular}{lllllcc}
      \hline
      \noalign{\smallskip}
     &   model & $N_{\rm H}$ & $\Gamma_{\rm x}$      
                            & $kT$ & $Z$ 
                            & $\chi^2_{\rm red}(d.o.f)$ \\
       \noalign{\smallskip}
      \hline
       \noalign{\smallskip}
       &       & 10$^{21}$ cm$^{-2}$ & &     
               keV & x solar & \\     
       \noalign{\smallskip}
      \hline
      \hline
      \noalign{\smallskip}
NGC\,383 & pl & 4.4 & $-5.1$ & -- & -- & 
1.38 (11)  \\
 & pl & 0.523$^{2}$ & $-1.9$ & -- & -- & 
3.25 (12)  \\
 & rs & 0.43 & -- & 1.14 & 0.35$^{3}$ & 
2.42 (11) \\
 & rs & 0.59 & -- & 0.98 & 0.10$^{3}$ & 
1.50 (11) \\
 & rs+pl & 0.69 & $-1.9^{3}$ & 0.55 & 0.35$^{3}$ &
1.12 (10) \\
 & rs+pl & 0.523$^{2}$ & $-1.9^{3}$ & 0.67 & 0.35$^{3}$ &
1.18 (11) \\
 & rs+rs & 0.523$^{2}$ & -- & 0.5/5$^{5}$ & 0.35$^{3}$ & 
 1.02 (11) \\
 & rs+rs & 0.523$^{2}$ & -- & 0.4/5$^{5}$ & 1.00$^{3}$ & 
 1.01 (11) \\
\noalign{\smallskip}
\hline
\noalign{\smallskip}
NGC\,380 & pl & 0.523$^{2}$ & $-1.8$ & -- & -- & 
5.58 (~7)  \\
 & rs & 0.523$^{2}$ & -- & 0.92$\pm{0.11}$ & 0.35$^{3}$ & 
0.84 (~7) \\
\noalign{\smallskip}
\hline
\noalign{\smallskip}
NGC\,379 & pl & 0.523$^{2}$ & $-2.2$ & -- & -- & 
3.38 (~4)  \\
 & rs & 0.523$^{2}$ & -- & 0.50$\pm{0.27}$ & 0.35$^{3}$ & 
1.03 (~4) \\
\noalign{\smallskip}
\hline
\hline
\noalign{\smallskip}
IGM & rs; total & 0.523$^{2}$ & -- & 1.46$\pm{0.08}$ & 0.35$^{3}$ & 
1.14 (22) \\
 & rs; total & 0.523$^{2}$ & -- & 1.56$\pm{0.12}$ & 1.00$^{3}$ & 
0.73 (22) \\
 & rs; inn. & 0.523$^{2}$ & -- & 1.48$\pm{0.12}$ & 0.35$^{3}$ & 
1.43 (18) \\
 & rs; mid. & 0.523$^{2}$ & -- & 1.56$\pm{0.17}$ & 0.35$^{3}$ & 
1.79 (19) \\
 & rs; out. & 0.523$^{2}$ & -- & 1.37$\pm{0.14}$ & 0.35$^{3}$ &
1.42 (14) \\
      \noalign{\smallskip}
      \hline
      \noalign{\smallskip}
  \end{tabular}

\noindent{\small  
$^{(2)}$ fixed to the Galactic value; $^{(3)}$ fixed; $^{(5)}$ temperature of second 
rs component, fixed \\
}
  \vspace{-0.4cm}
   \end{table*}

  \begin{figure}[t]      
      \vbox{\psfig{figure=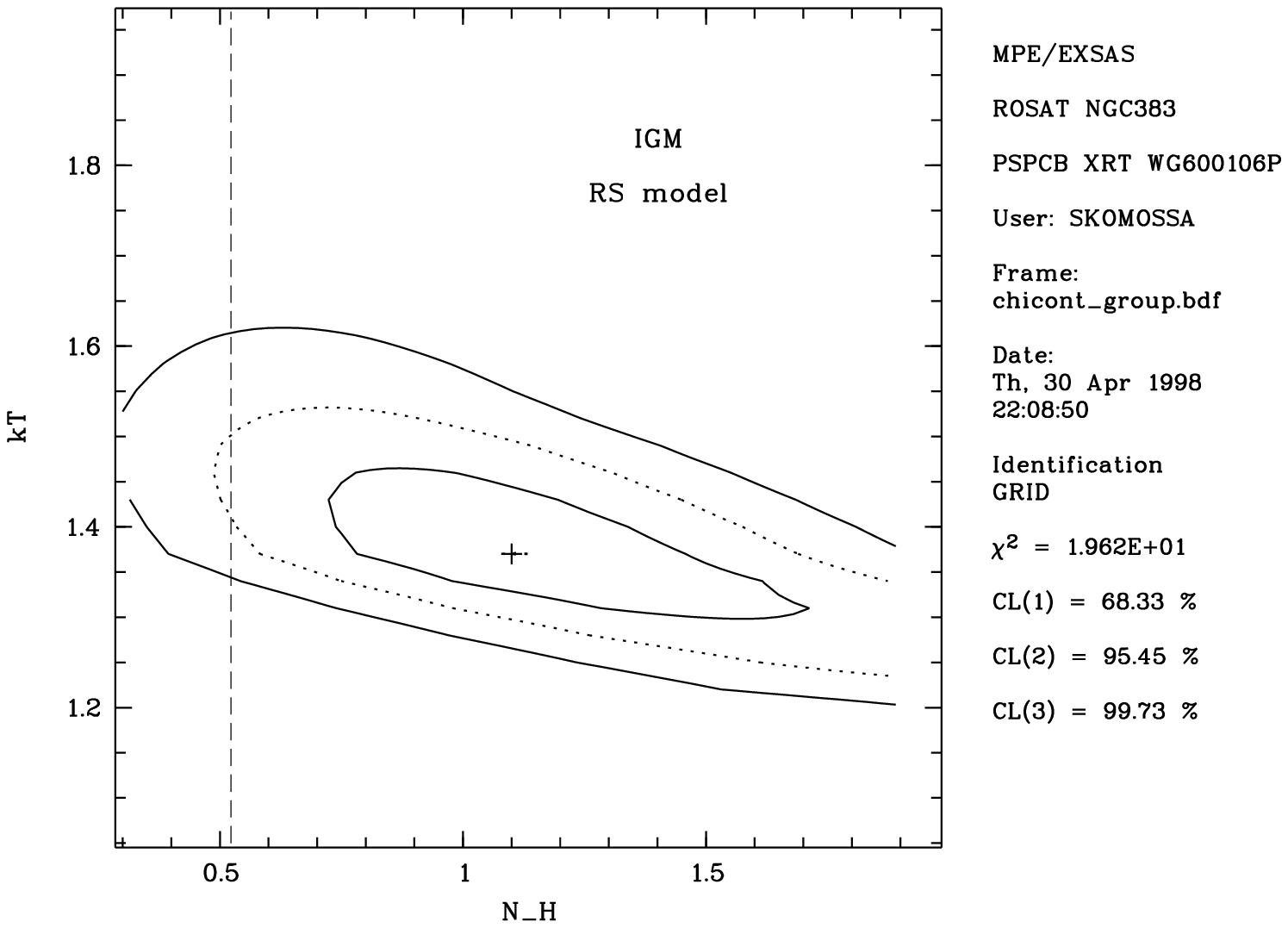,width=7.7cm,%
          bbllx=2.2cm,bblly=1.1cm,bburx=14.5cm,bbury=12.2cm,clip=}}\par
 \caption[chi]{Error contours in $kT$ (in units of keV), $N_{\rm H}$
(in 10$^{21}$ cm$^{-2}$)
 for the rs model of the IGM of the NGC\,383 group.
The contours are shown for confidence levels
of 68.3, 95.5 and 99.7\%. The dashed line marks the
Galactic absorbing column density towards the X-ray center of the group.
}
 \label{chi}
\end{figure}

  \begin{figure}[t]      
      \vbox{\psfig{figure=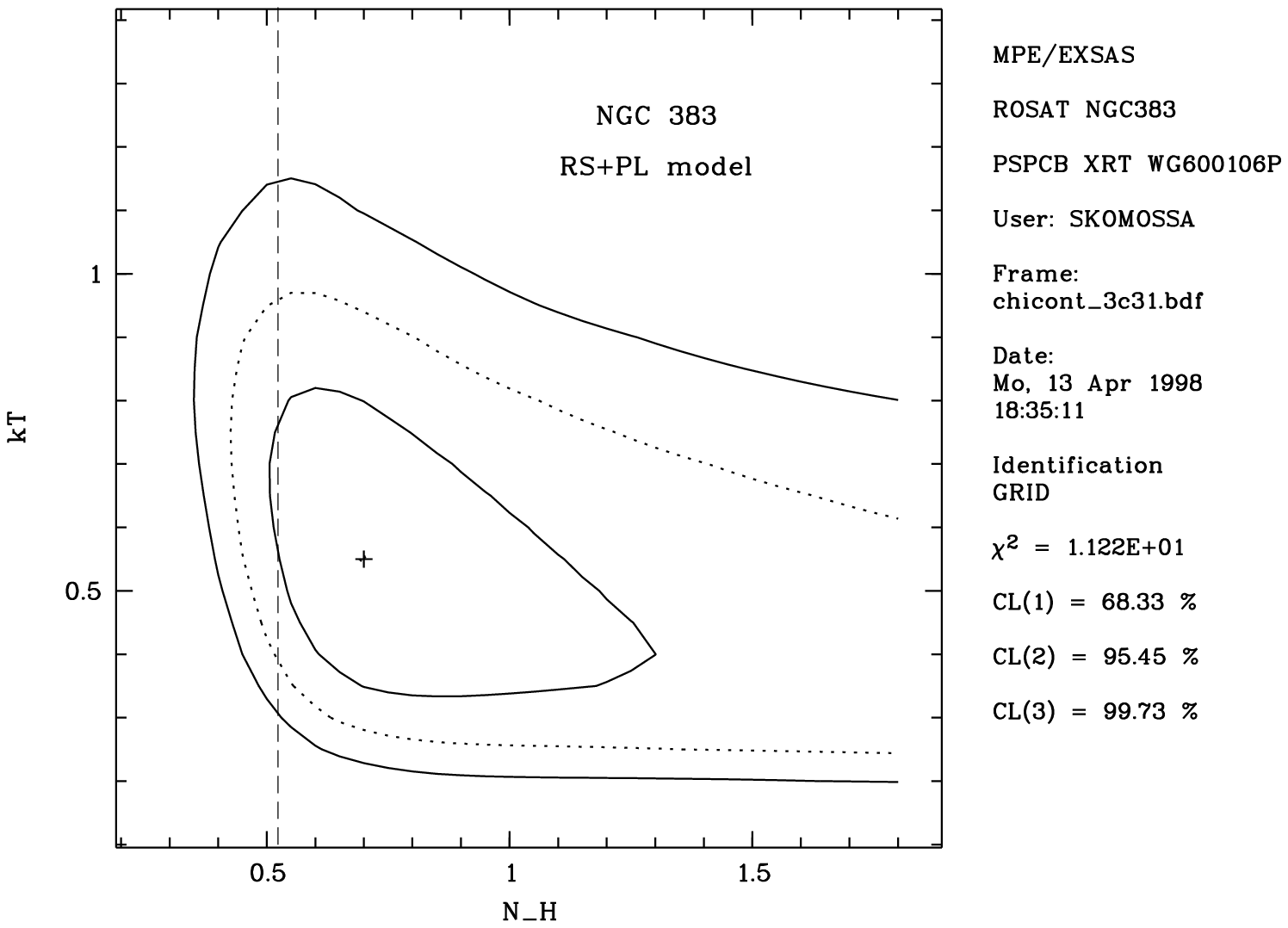,width=7.7cm,%
          bbllx=2.2cm,bblly=1.1cm,bburx=14.5cm,bbury=12.2cm,clip=}}\par
 \caption[chi2]{Error contours in $kT$ (in keV), $N_{\rm H}$ 
(in 10$^{21}$ cm$^{-2}$)
 for the rs+pl description of NGC\,383; the normalizations of the two components 
were free parameters, the pl index was fixed to --1.9. 
The contours are shown for confidence levels
of 68.3, 95.5 and 99.7\%. The dashed line marks the 
Galactic $N_{\rm H}$ towards NGC\,383. 
}
 \label{chi2}
\end{figure}

\subsubsection {Spatial analysis}

The X-ray emission from the direction of NGC\,383 is strongly peaked. 
To check how much of the emission might arise from a point source
we compared the radial source profile with the instrumental point-spread function (PSF).
We find that the emission from NGC\,383 
is not significantly extended beyond the
PSF of the PSPC. Thus, the data are consistent with 
the bulk of the X-ray emission arising from a point source. 

Performing a similar analysis for the HRI observation, we find a deviation
of the source profile from the HRI PSF. However, similar deviations are
found for the (presumably pointlike) F star which is located within the field
of view (after taking into account the appropriate off-axis PSF for the star).  
We conclude that in the present data there is no evidence for
source extent.  

\subsubsection {Temporal analysis}

\begin{figure}[h]
      \vbox{\psfig{figure=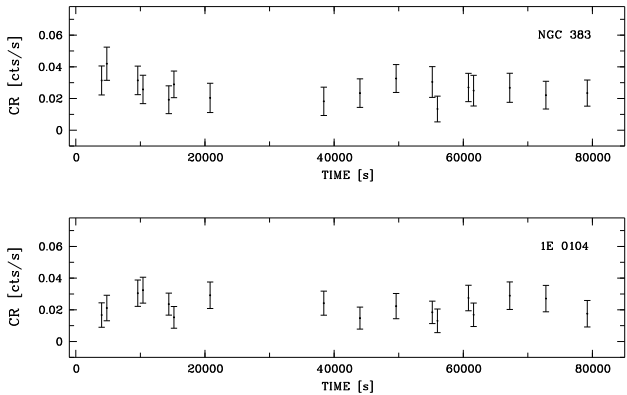,width=8.8cm,
          bbllx=3.0cm,bblly=2.5cm,bburx=18.0cm,bbury=12.2cm,clip=}}\par
 \caption[light]{PSPC X-ray lightcurve of NGC\,383 (upper panel) and 1E\,0104 (lower panel)
binned to time intervals of 800 s. The time
is measured in seconds from the start of the observation.
   }
 \label{light}
\end{figure}

The X-ray lightcurve of NGC\,383 is displayed in Fig. \ref{light}.
An AGN/point-source might reveal itself by variability (but not necessarily). 
We find a constant source flux within the errors.

\subsection {NGC\,379, NGC\,380, NGC\,384, NGC\,385}

A spectral analysis was performed for the two X-ray brightest galaxies,
NGC\,379 and NGC\,380. A pl model does
not provide an acceptable fit ($\chi{^{2}}_{\rm red}$ = 3.4 and 5.6) with residuals strongly 
indicative of the presence of an rs component. Such a model indeed gives an
excellent fit ($\chi{^{2}}_{\rm red}$ = 1.0 and 0.8). 
We find temperatures of 0.5 keV (NGC\,379) and 0.9 keV (NGC\,380).
The absorption-corrected fluxes for this model description in the
(0.1--2.4) keV band are $f_{\rm x} = 8.5\,10^{-14}$ erg/cm$^2$/s (NGC\,379)
and $f_{\rm x} = 1.85\,10^{-13}$ erg/cm$^2$/s (NGC\,380),
and the corresponding luminosities $L_{\rm x} = 1.1\,10^{41}$ erg/s (NGC\,379)
and $L_{\rm x} = 2.3\,10^{41}$ erg/s (NGC\,380). 

To estimate luminosities also for NGC\,384 and 385, which are too weak to allow
direct spectral fits, we adopted a rs spectrum of 0.5 keV. The derived luminosities
are given in Table 2.  

\begin{table} 
 \caption{Summary of the properties of the X-ray brightest group galaxies.}
 \label{fitres2}
 \begin{tabular}{ccccc}
 \hline
 \noalign{\smallskip}
\normalsize
galaxy & $CR$ & $kT^{(1)}$ & $L_{\rm (0.1-2.4) keV}$ & $L_{\rm B}$ \\  
  \noalign{\smallskip}
  \hline
  \noalign{\smallskip}
  & 10$^{-2}$ cts/s & keV & erg/s & erg/s \\
 \noalign{\smallskip}
 \hline
 \hline
 \noalign{\smallskip}
NGC\,379 & 0.58 & 0.5 & 1.1\,10$^{41}$ & 3.1\,10$^{43}$ \\ 
NGC\,380 & 1.24 & 0.9 & 2.3\,10$^{41}$ & 4.0\,10$^{43}$ \\ 
NGC\,383 & 2.40 & 0.6+pl & 4.7\,10$^{41}$ & 5.1\,10$^{43}$ \\ 
NGC\,384 & 0.21 & 0.5$^*$ & 5.0\,10$^{40}$ & 1.8\,10$^{43}$ \\ 
NGC\,385 & 0.41 & 0.5$^*$ & 7.6\,10$^{40}$ & 2.6\,10$^{43}$ \\ 
 \noalign{\smallskip}
 \hline
 \noalign{\smallskip}
 \end{tabular}

\noindent{\small $^{(1)}$ either determined directly from spectral fit (Sect. 4.1.1, 4.2)
or fixed to 0.5 keV (marked with `$^*$') in cases where the 
spectrum could not be fit directly.
}
   \end{table}

\subsection {$L_{\rm x} - L_{\rm B}$ relation} 

To compare the derived X-ray luminosities with blue luminosities
we used the observed blue magnitudes of de Vaucouleur et al. (1991 via NED;
see also Smith et al. 1997). 
For the extinction correction we converted the Galactic $N_{\rm H}$ as
given in Dickey \& Lockman (1990) into $A_{\rm B}$ assuming a standard gas/dust
ratio, the relation of Bohlin et al. (1978; see also Predehl \& Schmitt 1995), 
and the extinction curve as 
given in Osterbrock et al. (1989, his Tab. 7.2). 
This yields $A_{\rm B}$ = 0.38$^{\rm m}$. 
$L_{\rm B}$ was then calculated using $L_{\rm B} = 4\pi d^2 10^{(-0.4m_{\rm B}-5.19)}$
and assuming the same $z$=0.017 for all galaxies. 
Results are listed in Table 2.

\section{Discussion}

\subsection{The group}

\subsubsection{IGM}

The NGC\,383 group turns out to
be the brightest group in X-ray luminosity when compared to the samples
of Ponman et al. (1996) and Mulchaey \& Zabludoff (1998). 

How does it fit into the known $L_{\rm x}-T$ relation 
for groups and clusters
of galaxies (e.g., Fabian et al. 1994, White 1996, Ponman et al. 1996, 
Arnaud \& Evrard 1998, Reiprich 1998)?  
Since $T$ is outside the range for which Ponman et al. find a very steep $L_{\rm x}-T$ 
dependence, 
we use the relation of Markevitch (1998) which predicts
a (0.1-2.4 keV) X-ray luminosity $L_{\rm x} \simeq 3\,10^{43}$ erg/s.
This agrees well with the observed value of $1.5\,10^{43}$ erg/s.

For the given $kT$ = 1.5 keV, and a galaxy velocity dispersion of $\sigma$ = 466 km/s
(Ledlow et al. 1996; see also Sakai et al. 1994) we derive 
$\beta_{\rm spec} = {\mu m_{\rm p} \sigma_{\rm r}^2 \over {k T}} = 0.95$.  

The low value of the slope parameter $\beta$ derived from the X-ray spatial analysis
is in line with earlier findings of a trend of
decreasing $\beta$ toward lower $T$ (e.g., David et al. 1990, White 1991,
Arnaud \& Evrard 1998)   
which is reproduced by models of cluster formation that incorporate galactic
winds (e.g., Metzler \& Evrard 1997).

Assuming spherical symmetry and isothermality, the
total gravitating mass within 1 Mpc amounts to $M_{\rm total}$ = 0.6\,10$^{14}$ M$_{\odot}$.
We find a gas mass fraction of 21\% which is at the upper end of the values typically
observed in groups (3--25\%, e.g., David et al. 1995, B\"ohringer 1995) but 
not inconsistent with similar results for such X-ray luminous
groups as for example NGC\,533 or NGC\,4104 (Mulchaey et al. 1996).  
The gas mass fraction slightly decreases with
decreasing radius.

\begin{table}             
 \caption{Summary of the properties of the NGC\,383 group of galaxies as derived from
the X-ray analysis.}
\normalsize
     \label{gprop}
      \begin{tabular}{l}
      \hline
      \noalign{\smallskip}
spectral fits: \\
\indent ~~ $kT$=1.5 keV, $L_{\rm x}^{\rm 0.1-2.4 keV}$ = 1.5\,10$^{43}$ erg/s\\
beta-model results:  \\
\indent ~~ $S_{\rm 0} = 2.7\,10^{-3}$ cts/s/arcmin$^2$, $\beta = 0.38$,
$r_{\rm c} = 73$ kpc \\
central density, mass: \\
\indent ~~ $n_0$ = 1.3\,10$^{-3}$ cm$^{-3}$; $M_{\rm total}$ = 0.6\,10$^{14}$ M$_{\odot}$, \\
\indent ~~ gas mass fraction 21\% (at $r$ = 1\,Mpc) \\
      \noalign{\smallskip}
      \hline
 \end{tabular}
   \end{table}

Concerning the morphology of the IGM, we note that 
the rather spherically symmetric 
shape of the extended emission, as compared to the completely different morphology
defined by the bright ellipticals aligned in a chain, 
argues against an origin of the gas in terms of halos of the chain galaxies,
but rather for an association with the global group potential.   

The X-ray emission of the IGM can be traced out
to about 1 Mpc, which turns out to be about
the virial radius of the group if we use the
total mass value determined from the X-ray observations
and the assumption that the virial radius is approximately
characterized by the region inside which the
mean overdensity is a factor of 200 above the critical
density of the universe (e.g., Evrard et al. 1996).

The fairly spherically symmetric appearance of the group's X-ray halo
(except possibly for some faint outer
extensions)
taken together with
the perfect consistency of the mass determined
from the velocity dispersion and the X-ray properties,
and the findings of Ledlow et al. (1996) that
the galaxy velocity distribution does not significantly
deviate from a Gaussian, 
implies that the matter in the group inside a radius
of about 1 Mpc is most probably quite relaxed.

The position of the central galaxy NGC\,383 is found to be slightly off-set from
the center of the extended X-ray emission, and thus presumably from the center of
the dark matter potential. 
Such off-sets of cD galaxies have also been observed in a number of other poor systems 
(e.g., AWM\,7, Neumann \& B\"ohringer 1995; Fornax cluster, 
Ikebe et al. 1996) and some Abell clusters (e.g., Lazzati \& Chincarini 1998). 
In Lazzati \& Chincarini (1998), this is traced back to a small-amplitude  
oscillation of the cD galaxy around the bottom of the cluster potential.  

\subsubsection{Presence of a cooling flow ?}

The cooling time in the center is $t \simeq 2.7\,10^{10}$ yr, i.e. no `large-scale'
cooling flow is expected to have developed. 
Further, the enhanced X-ray emission from the direction
of NGC\,383 is consistent with originating from a point source. 

We also note that although strong low-ionization optical emission lines have been reported
for some central galaxies in cluster cooling flows (e.g., Cowie et al. 1983, Heckman 1989,
Crawford \& Fabian 1992),
the morphology of the ring- or disk-like emission line region in NGC\,383 (Owen et al. 1990,
Fraix-Burnet et al. 1991) does not argue for
a connection to a cooling flow. 
Further, we find the locus of NGC\,383 in the emission
line-ratio diagram [SII] vs. [NII] to ly outside of 
the `class I' and `class II' cooling flow nebulae of 
Heckman et al. (1989; see their Fig. 6).
(This alone does not exclude the presence of a cooling flow, though, since
not all of them are associated with emission line nebulae.)

\subsubsection{Radio - X - relations, pressure estimate}

  \begin{figure}[t]      
      \vbox{\psfig{figure=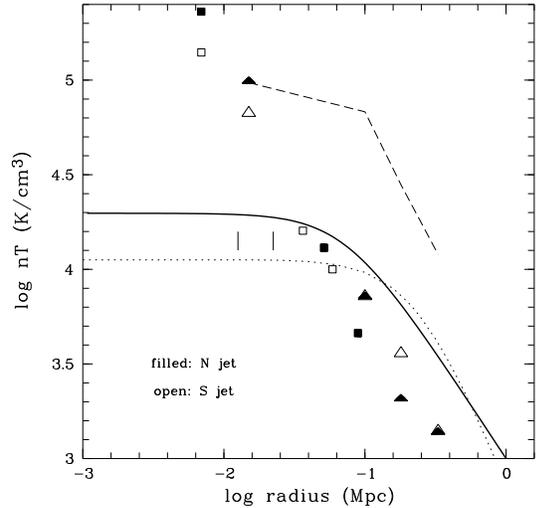,width=7.7cm,%
          bbllx=2.5cm,bblly=1.1cm,bburx=15cm,bbury=12.2cm,clip=}}\par
 \caption[pressure]{ Comparison of different pressure estimates.
 The solid line gives the thermal pressure as derived from
the present \ros X-ray observation. The symbols mark the non-thermal pressure
as given in Strom et al. (1983; squares) and Morganti et al. (1988, triangles);
the filled symbols are for the northern jet, the open ones for the southern jet.
The vertical bars mark a scale of 25\arcsec~ (left) and the optical extent of
NGC\,383 (as given in NED; right).
Also drawn is the change in thermal pressure according to two previous
estimates (dashed line: Morganti et al. 1988; dotted: T97).
}
 \label{pressure}
\end{figure}

Although 3C\,31 is only a moderately bright radio galaxy, its jets
could be studied in detail due to its proximity.
Concerning the origin, morphology and confinement of the jets, several models
were explored (e.g., Blandford \& Icke 1978, Bridle et al. 1994, Bicknell 1994).   

The gas density and temperature derived for the X-ray gas allow a comparison
with the pressure of the radio gas, and an assessment of the confinement
of the jet material.  
In Fig. \ref{pressure} we compare the pressure of the radio emitting region 
as given in Strom et al. (1983) and Morganti et al. (1988){\footnote{To derive
the equipartition pressure, 
they used the equations of Pacholczyk (1970) and made the standard assumptions of
equal energy density in protons and electrons, filling factor of 1, and
a powerlaw representation of the radio spectrum with cut-offs at 10 MHz and
100 GHz.}} with the thermal pressure of the X-ray gas derived from the run of 
density (Sect. 3.3) and a temperature of $kT$=1.5 keV.
Whereas in the central region (i.e. within NGC\,383), there seems to be a strong
overpressure of the radio gas (but note that there certainly is an additional
contribution 
to thermal pressure from
higher-density gas within the galaxy), pressure equilibrium is reached
at about 35 kpc (projected distance from the center). Further out the thermal pressure 
increasingly exceeds the nonthermal pressure.

It is interesting to note that Bridle et al. (1980) 
find the expansion rate of the jets of 3C\,31 transverse to their length
to decrease with increasing distance from the radio core.
This may be related to the relative increase of the thermal pressure
of the ambient medium  
with increasing radius.
While Bridle's trend refers mainly to the presently in X-rays barely resolved
core region it may be interesting to explore this relation further with 
higher-resolution X-ray data.

A similar comparison of
thermal vs. non-thermal pressure was performed in
T97. They derived a somewhat different surface brightness profile
and thus change in thermal pressure with the consequence that the radius 
where both pressure values are of the same order shifts further out
which led them to suggest that NGC\,383 might have a giant halo that
escaped detection in the (short) HRI exposure.

\subsection{Individual sources}

\subsubsection{ $L_{\rm X} - L_{\rm B}$ relation} 

The galaxies of the chain are among those with high $L_{\rm x}/L_{\rm B}$
and show a larger spread in $L_{\rm x}$ than in $L_{\rm B}$ as often observed
(e.g., Eskridge et al. 1995). 

The high $L_{\rm x}$ in cluster/group ellipticals (e.g., Fig. 1 of Brown \& Bregman 1998,
Fig. 2 of Beuing et al. 1998, Fig. 1 of Irwin \& Sarazin 1998) are usually traced back to the 
influence of the surroundings, via accretion of gas from the group
environment (e.g., Beuing et al. 1998), or stiffling of winds by the
ambient medium (Brown \& Bregman 1998); see also Mathews \& Brighenti (1998). 
The low $L_{\rm X}/L_{\rm B}$ systems
seem to be dominated by discrete sources, mainly LMXBs (e.g., Canizares et al. 1987,
Irwin \& Sarazin 1998).   

\begin{figure}[ht]      
      \vbox{\psfig{figure=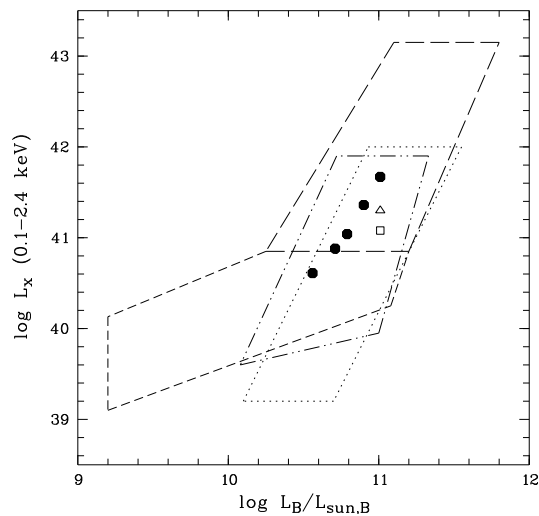,width=7.7cm,%
          bbllx=2.5cm,bblly=1.1cm,bburx=15cm,bbury=12.2cm,clip=}}\par
 \caption[lxlb]{Locations of the chain galaxies in the $L_{\rm x}-L_{\rm B}$ 
diagram (fat circles; NGC\,383, NGC\,380, NGC\,379, NGC\,385, NGC\,384
from top to bottom; see also T97). For comparison, we sketch
the regions populated by some larger samples of ellipticals (note that these $L_{\rm x}$
were derived under slightly different assumptions, i.e. source extraction
radii and energy bands which we did not correct
for since we only intend to show the rough trends). 
Dotted: Brown \& Bregman 1998, dot-dashed: Canizares et al. 1987, 
long-dashed: Beuing et al. 1998 (detections), short-dashed: Beuing et al. 1998 (upper
limits). The open square (triangle) marks the locus
of NGC\,383 after subtraction of the contribution of the pl (hot rs) component to $L_{\rm x}$.  
} 
 \label{lxlb}
\end{figure}

The present galaxies show a wide range in $L_{\rm x}$ reaching up 
 to the high end of the $L_{\rm x}$/$L_{\rm B}$ relation. 
(Fig. \ref{lxlb}; note that the objects of very high $L_{\rm x}$
of the sample of Beuing et al. 1998 are all ellipticals in 
the {\em cluster} environment).
The X-ray luminosities generally exceed those expected from a discrete source contribution
and the X-ray properties
of most galaxies are consistent with the bulk of the X-ray emission arising from the ISM of the 
individual galaxies.  
NGC\,383 is separately discussed in the next Section. 

\subsubsection{NGC\,383}

The X-ray analysis suggests the presence of two components
in the X-ray spectrum of NGC\,383: emission from a rs plasma of $kT \simeq$ 0.4--0.7 keV
and a second component that was parameterized as powerlaw or second rs contribution.

Compact pl emission could originate from an AGN
or via SSC at the base
of the radio jet; e.g., Jones et al. (1974), Marscher (1987), Birkinshaw \& Worrall (1993). 
Using the correlation of 1\,keV X-ray core flux with 5\,GHz radio core flux 
of Worrall et al. (1994, their Fig. 3; see also Fabbiano et al. 1984) 
we find 3C\,31 above this relation (a factor $\sim$6 in X-ray flux 
for the given radio core flux{\footnote{we used the 5\,GHz core radio flux 
as given in Lara et al. (1997), 90 mJy, 
and the rs+pl X-ray spectral model with $N_{\rm H}=N_{\rm Gal}$ for which
we derive a 1\,keV X-ray flux in the pl component of 
0.058 $\mu$Jy}}), and within the region populated by lobe-dominated quasars.
   
Is there evidence for a relatively {\em unobscured} view on an AGN ?
Optical emission line ratios were presented by Owen et al. (1990),
who found the line ratios in the disk to be similar to those
at the nucleus ([NII]/H$\alpha$=1.5, [SII]/H$\alpha$=0.5, [OIII]/H$\alpha$=0.1)
and concluded that the disk ionization is probably driven
by the nucleus.
Plotting these line ratios in the diagnostic diagrams of Veilleux \& Osterbrock (1987)
we find them to be LINER-like.  
The dominant excitation mechanism in LINERS is
still under discussion, 
but there is growing evidence that the bulk of them is accretion-powered (e.g., 
Ho et al. 1998). 
Owen et al. also briefly mention the presence of a broad emission line component.

Alternatively, two-component rs models fit the X-ray spectrum; 
a second component may originate
from non-perfect correction of the cluster contribution 
or a temperature range in emission
from the ISM of the galaxy. In fact, more than one temperature component  
is favoured for nearly every early-type galaxy examined in, e.g.,  
Matsushita et al. (1994), Matsumoto et al. (1997) and 
Buote \& Fabian (1998).
Whereas the hot component is occasionally interpreted in terms of emission from
discrete sources (e.g., Matsushita et al. 1994), we find its luminosity
($L_{\rm x,pl} = 3.5\,10^{41}$ erg/s, $L_{\rm x,rs} \simeq 3\,10^{41}$ erg/s
if parameterized by a pl or rs model, respectively)
to be higher than expected from discrete sources 
(e.g., Canizares et al. 1987) for the given $L_{\rm B}$. 

In conclusion, although the exact shape and strength of the hard 
X-ray component cannot be determined with present data, its 
presence and high luminosity suggest as origin emission 
from an active nucleus in NGC\,383.

\section{1E\,0104+3153 ($z$=2 BAL quasar/ elliptical/ IGM of group)} 

The source 1E\,0104 was serendipitously detected as bright X-ray emitter by \ein~
and was found to be close to a 
BAL quasar with $z$=2.027 (Stocke et al. 1984). 
Since the QSO is also located very near (10\arcsec) to a  giant elliptical
galaxy at $z$=0.111, which is part of a small group,
it remained unclear from which of the three the X-ray emission
originated: the QSO, the giant elliptical, or the IGM
of the small group (Stocke et al. 1984). Gioia et al. (1986), in a deep \exo observation, 
could not detect X-ray emission in the 0.05--2 keV band, which
they traced back to either variability (intrinsic to
the source, or caused by a microlensing event during the \ein~
observation; which would favour the quasar identification
of the X-ray source), or a strongly absorbed X-ray spectrum. 
The incidental \ros PSPC observation of the source was briefly discussed by 
Ciliegi \& Maccacaro (1996) in a large sample
study of \ros spectra of EMSS AGN. 
They applied a powerlaw to the spectrum and found excess absorption. 

Below, we analyze the X-ray spectral properties in more detail and also perform
a timing analysis, in order to get clues on the origin of the X-ray emission
(elliptical galaxy, group or 
quasar).  

\subsection {Data analysis}

\subsubsection {Spatial analysis}
\begin{figure}[thbp]
      \vbox{\psfig{figure=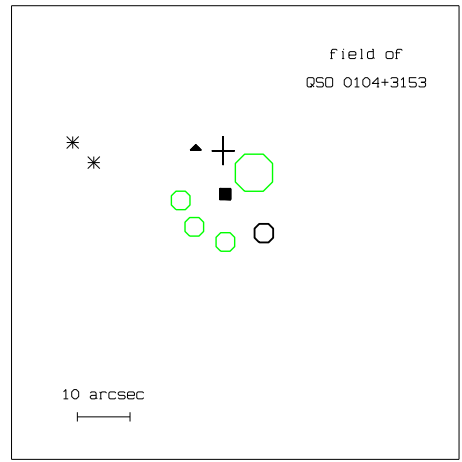,width=8.8cm,%
          bbllx=2.2cm,bblly=1.1cm,bburx=14.5cm,bbury=11.2cm,clip=}}\par
 \caption[pos]{Optical and X-ray positions for the field around QSO 0104 and
the nearby elliptical galaxy.
The circles and asterisks correspond to optical positions derived by Stocke et al. (1984;
large circle: giant elliptical, small fat
circle: QSO\,0104, further circles: galaxies, members of a small group, asterisks: stars).
The filled symbols and the cross mark X-ray positions measured by different instruments
(cross: \ein~IPC, filled triangle: \ros PSPC, filled square: \ros HRI).
}
 \label{pos}
\end{figure}

The HRI data improve the position of 1E\,0104, but not much, since the source is weak and located
near the border of the fov. In Fig. \ref{pos} we compare the \ros HRI position
with the optical positions of the quasar and the galaxies
(as given in Stocke et al. 1984; all coordinates
were converted to J\,2000) and the \ein~IPC position.
Both sources, the quasar and the elliptical, remain within the error circle.

We note, however, that the X-ray source appears to be extended.
Although the analysis is complicated by a newly detected closeby second source, a
comparison with the X-ray emission from the F star at nearly the same
off-axis angle, clearly suggests that the X-ray emission from 1E\,0104
does not originate from a point source (see Fig. 1; the F star is the brightest
source at the right border of the image, 1E\,0104 is the one at the lower border).

\subsubsection {Temporal analysis}
We do not detect short-time variability during the PSPC
observation (Fig. \ref{light}). To examine the long-term trend,
we converted the PSPC countrate to HRI countrate assuming
constant spectral shape. Again, the
source emission is found to be constant.
We then compared
with the \ein~ IPC flux (this obs. was performed 11 yrs prior to the
\ros PSPC observation) given in Stocke et al. (1984). To this end, we first
re-fit the \ros data with the same spectral model as in Stocke et al.
(they used \G=--1.5 and $N_{\rm H} = 0.3\,10^{20}$ cm$^{-2}$; see Gioia et al. 1986
for similar parameters). We then derive an observed flux (converted to the 0.3 -- 3.5 keV band)
of $f_{\rm ros} = 3.96\,10^{-13}$ erg/cm$^2$/s which is in excellent agreement
with the \ein~ flux of $f_{\rm ein} = 4.0(\pm{0.7})\,10^{-13}$ erg/cm$^2$/s.

\subsubsection{Spectral analysis}
A powerlaw with absorption fixed to the Galactic value does not give a good fit
with \G = --1.7 and $\chi{^{2}}_{\rm red}$ = 1.5. The fit improves after allowing for excess
cold absorption. This yields $N_{\rm H} = 2.9\,10^{21}$ cm$^{-2}$,
\G = --3.2 and $\chi{^{2}}_{\rm red}$ = 1.1.
Alternatively, the spectrum can be successfully described in terms of
a rs model.
In this case we used $z$=0.111 since rs emission is more likely to originate
from the elliptical and/or IGM.
If $N_{\rm H}$ is treated as free parameter it is consistent with $N_{\rm gal}$
but comes with large errors. We obtain $N_{\rm H} = (0.66^{+0.34}_{-0.16})\,10^{21}$ cm$^{-2}$
and $kT = 1.89^{+1.11}_{-0.29}$ keV ($\chi{^{2}}_{\rm red}$ = 1.0).
For $N_{\rm H}$=$N_{\rm gal}$, $kT$ = 1.9 keV and $L_{\rm x} = 3\,10^{43}$ erg/s.

Plugging this temperature into the $L_{\rm x} - T$ relation for groups/clusters
(Markevitch 1998) we predict a (0.1--2.4 keV) X-ray luminosity of 5\,10$^{43}$ erg/s
which is of the order of the observed value and thus consistent with an identification
of 1E\,0104 with the IGM of the small group.

  \begin{figure} 
      \vbox{\psfig{figure=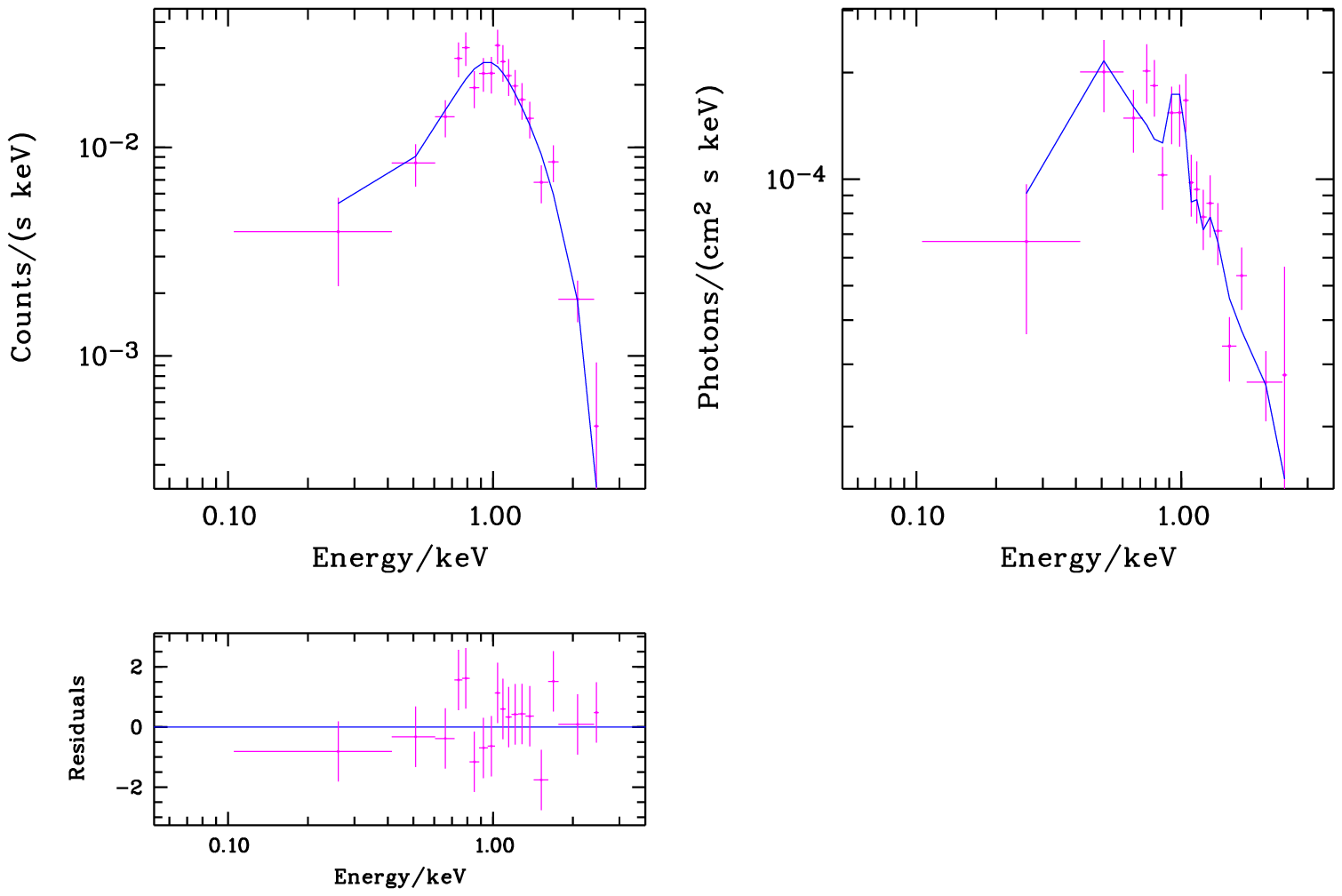,width=7.3cm,%
          bbllx=2.5cm,bblly=1.1cm,bburx=10.1cm,bbury=12cm,clip=}}\par
\caption[SEDe]{Observed (crosses) and fit (solid line) X-ray spectrum (upper panel)
and residuals (lower panel) for the Raymond-Smith model fit to 1E\,0104.
}
\label{SEDe}
\end{figure}

\subsubsection {Discussion}

The X-ray emission of 1E\,0104 turned out to be constant.
In particular, there is excellent agreement with the flux derived for the \ein~
observation (adopting the same spectral model). This renders one of several
possibilities discussed in Stocke et al. (1984) -- the one that the quasar
was temporarily brightened via microlensing during the \ein~ observation -- unlikely.

We confirm the excess cold absorption in case of the pl description
of the X-ray spectrum, not unexpected for
a BAL quasar,  but find an rs model to be similarly successful.
The best-fit parameters of the latter model, which due to its shape would argue
for an origin from the group's IGM or brightest group elliptical, are well consistent
with the known $L_{\rm x}$-$T$ relation of clusters and groups of galaxies, i.e.,
with the IGM as counterpart. 
Further, the derived $L_{\rm x} \simeq 3\,10^{43}$ erg/s would be rather high 
for an elliptical outside a rich cluster environment
(particularly for the observed (Stocke et al. 1984) 
R-magnitude of $m_{\rm R}$ = 15.1), but usual for the IGM.  
The HRI position is intermediate between the elliptical and the quasar,
and consistent with all three identifications. However, there is evidence
that the source emission is extended.  

In summary, if only one emitter dominates the X-ray flux (instead of a contribution from {\em all three}
potential counterparts, which would strongly complicate matters but cannot be excluded)
it seems that the available data (constancy in source flux, high $L_{\rm x}$, fit into
$L_{\rm x}$-$T$ relation, evidence for source extent) 
favour an identification with the IGM of the nearby
group at $z$=0.111.
A deep high spatial resolution observation {\em centered} on 1E\,0104 should finally
resolve the counterpart question. 

\section {Summarizing conclusions}

We have presented a study of the X-ray properties of the NGC\,383 group of
galaxies, extending the work of T97, and of the source 1E\,0104+3135. 

The properties of the intra-group medium derived from the
\ros PSPC observation are summarized in Tab. 3.
The X-ray emission of the IGM can be traced out
to about $1 h_{50}^{-1}$ Mpc, which turns out to be about
the virial radius of the group.
Several lines of evidence were presented that 
the group inside this radius is quite relaxed.   
With the given depth of the \ros PSPC observations
we can therefore characterize the entire galaxy system
as far as it has approached a dynamical equilibrium state.
For this part of the group we find a total mass
of $6\,10^{13} h_{50}^{-1}$ M$_{\odot}$ and a gas mass
fraction of 21\%. The latter value is at the upper
end of the distribution for groups.
The result implies that the gravitational potential
of the group is deep enough to prevent a majour
gas loss, in contrast to less massive groups (Davis et al. 1998).

The surface brightness profile
is characterized by a slope parameter $\beta \simeq 0.4$
which is shallower than for most of the galaxy clusters
but a quite common value for groups. The temperature
of $1.5$ keV found for the IGM is well consistent
with the $L_x - T$ relation giving further support
to the picture that the group is a well relaxed
and normal system. 

With an estimated central cooling time larger than
the Hubble time no central cooling flow is expected
and no signature for it is found neither in the
spatial and spectral X-ray data nor in the optical
spectrum.

We do not find any conspicuous spatial correlation of X-ray 
emission and radio jet which might be partly due to the
narrowness of the jet and the 2D view of the 3D source structure.  

We also    
discussed the X-ray properties of the well-studied
radio galaxy 3C\,31 which is located near the center of the extended
X-ray emission.   
If one wishes to avoid excessively depleted metal abundances, 
the spectrum of 3C\,31 is best described by a two-component model,
consisting of a low-temperature rs component and a hard tail (pl or second rs),
confirming T97.
The soft component contributes with $\simeq3\,10^{41}$ erg/s to the total X-ray luminosity of
$L_{\rm x} \simeq5\,10^{41}$ erg/s
(assuming metal abundances of 0.35$\times$solar). 
No temporal variability in the X-ray flux is detected. 

X-ray emission from the direction of the interesting source 1E\,0104+3135, by chance 
located in the field of view, was analyzed. 
One scenario 
(transient brightening of the $z$=2 BAL QSO\,0104+3135 due
to lensing during the earlier \ein~observation) 
turned out to be unlikely. 
Besides an absorbed powerlaw, the spectrum of 1E\,0104 can be described by a Raymond-Smith
model with $kT \simeq$ 2 keV resulting in an intrinsic luminosity
of $L_{\rm x} \simeq 3\,10^{43}$ erg/s at $z$=0.111.  
Although no potential counterpart (QSO, nearby elliptical galaxy 
or IGM of the small group to which the elliptical belongs) can be safely
ruled out at present, there are several hints 
(constancy in source flux, high $L_{\rm x}$, consistency with 
$L_{\rm x}$-$T$ relation for groups/clusters, evidence for source extent)
for an identification of the X-ray source 
with the IGM of the nearby group of galaxies at $z$=0.111. 

\begin{acknowledgements}
We acknowledge support from the Verbundforschung under grant No. 50\,OR\,93065.
It is a pleasure to thank Vadim Burwitz for help with Figure 4. 
The \ros project is supported by the German Bundes\-mini\-ste\-rium
f\"ur Bildung, Wissenschaft, Forschung und Technologie 
(BMBF/DLR) and the Max-Planck-Society.
This research has made use of the NASA/IPAC extragalactic database (NED)
which is operated by the Jet Propulsion Laboratory, Caltech,
under contract with the National Aeronautics and Space
Administration.
\end{acknowledgements}

\end{document}